\documentclass[onecolumn]{aa} 
\usepackage{graphicx}
\usepackage{txfonts}
\begin{document}
   \title{Onset mechanism of an inverted U-shaped solar filament eruption revealed by NVST, SDO, and STEREO-A observations}

   \author{Jincheng Wang\inst{1,2} 
   \and Xiaoli Yan\inst{1,2} 
   \and Qiangwei Cai\inst{3}
   \and Zhike Xue\inst{1,2} 
   \and Liheng Yang\inst{1,2}
   \and Qiaoling Li\inst{4}
   \and Zhe Xu\inst{1,2}
   \and Yunfang Cai\inst{1,2}
   \and Liping Yang\inst{1,5}
   \and Yang Peng\inst{1,5}
   \and Xia Sun\inst{1,6}
   \and Xinsheng Zhang\inst{1,5}
   \and Yian Zhou\inst{1}
    }
   \institute{Yunnan Observatories, Chinese Academy of Sciences, Kunming Yunnan 650216, PR China\\
         \email{wangjincheng@ynao.ac.cn}
         \and
             Yunnan Key Laboratory of Solar Physics and Space Science, Kunming 650011, PR China
            \and
            Institute of Space Physics, Luoyang Normal University, Luoyang, 471934, PR China
            \and
             Department of Physics, Yunnan University, Kunming 650091, PR China
            \and
            University of Chinese Academy of Sciences, Yuquan Road, Shijingshan Block Beijing 100049, PR China
            \and
            School of Physics and Electronic Information Technology, Yunnan Normal University, Kunming 650500, PR China          
             }

   \date{Received **** **, ****; accepted **** **, ****}
   
   \abstract
   {Solar filaments, also called solar prominences when appearing on the solar limb, consist of dense, cool plasma suspended in the hot and tenuous corona, which are the main potential sources of solar storms.}
   {To understand the onset mechanism of solar filaments, we investigate the eruption process of an inverted U-shaped solar filament and two precursory jet-like activities.}
   {Utilizing observations from the New Vacuum Solar Telescope (NVST), Solar Dynamics Observatory (SDO), and Solar Terrestrial Relations Observatory-Ahead (STEREO-A), we investigate the event from two distinct observational perspectives: on the solar disk using NVST and SDO, and on the solar limb using STEREO-A. We employ both a non-linear force-free field model and a potential field model to reconstruct the coronal magnetic field, aiming to understand its magnetic properties.}
   {Two precursor jet-like activities were observed before the eruption, displaying an untwisted rotation. The second activity released an estimated twist of over two turns. During these two jet-like activities, ``Y''-shaped brightenings, newly emerging magnetic flux accompanied by magnetic cancellation, and the formation of newly moving fibrils were identified. Combining these observational features, it can be inferred that these two precursor jet-like activities released the magnetic field constraining the filament and were caused by newly emerging magnetic flux. Before the filament eruption, it was observed that some moving flows had been ejected from the site as the onset of two jet-like activities, indicating the same physical process as two jet-like activities. Extrapolations revealed that the filament laid under the height of the decay index of 1.0 and had strong magnetic field (540 Gauss) and a high twist number (2.4 turns) before the eruption. An apparent rotational motion was observed during the filament eruption.
}
    {We deduce that the solar filament, exhibiting an inverted U-shape, is a significantly twisted flux rope. The eruption of the filament was initiated by the release of constraining magnetic fields through continuous magnetic reconnection. This reconnection process was caused by the emergence of newly magnetic flux.}
    
 \keywords{Solar filament --
                Solar filament eruptions  --
                Solar activity
               }         
 \titlerunning{Onset mechanism of an inverted U-shaped solar filament revealed by NVST, SDO, and STEREO-A observations}
 \authorrunning{Wang et al.}
 \maketitle

 \section{Introduction}
 Solar filaments, ones of the most fascinating structures in the Sun, consist of cool and dense materials suspended in the hot and tenuous solar corona, appearing as dark elongated features on the solar disk. They exhibit as bright cloud-like structures when observed at the solar limb, also named solar prominences \citep[e.g.,][]{mar98,mac10,wan22a}. They often lie above the magnetic polarity inversion lines (PILs) separating the positive and negative polarities of the photospheric magnetic field \citep{bab55,mar98}. According to their locations, they can be divided into three classes: active region filament, intermediate filament, and quiescent filament. Magnetic fields play a key role in the stability, formation, and eruption of filaments in the corona \citep{mac10}. The eruption of filaments would yield solar storms (such as solar flares, global waves, and coronal mass ejections (CMEs)), which have a significant effect on the solar atmosphere and interplanetary space \citep[e.g.,][]{wan20,zho21,wan22b}. 

Understanding the filaments/prominences in the corona is the main subject for solar physics, including their formations, eruptions, and magnetic structures \citep{mar98,che20}. Magnetic flux rope and magnetic dips had been often discovered in many observations of the filaments, which suggests that filament materials are supported by the magnetic tension \citep{gil01,su15}. Photospheric flux cancellation and convergence are thought to be crucial to the formation of filament magnetic structures \citep[e.g.,][]{cha01,wan07,yan16}. It is widely accepted that the magnetic structures of filaments are generally formed through surface actions involving magnetic reconnections, rather than subsurface actions \citep{van89,wan17}. \cite{yan15} suggested that sunspot rotations and shearing motions are responsible for the formation of two homologous filaments. On the other hand, solar jets or some small-scale activity around the filament might be an efficient way to carry the materials for the filament \citep[e.g.,][]{wan18,wan19,li2023}. 

Based on numerous numerical simulations and observations, many candidate mechanisms had been proposed to explain how the solar filaments erupt, including reconnection driven processes (such as breakout models \citep{ant99}, tether-cutting models \citep{mor01}, and flux emergence models \citep{che00,lin01}) and ideal MHD instabilities (such as kink instability \citep{sak76,hoo79}, Torus instability \citep{kli06}, and catastrophic model \citep{lin00}). In the first category, the initiation mechanism relies on magnetic reconnections that rearrange the magnetic field, leading to the destabilization of the system. Magnetic reconnections beneath or above the filaments play a crucial role in triggering the eruption of the filament \citep{che18,lea22}. Newly emerging magnetic flux can also be considered to be a trigger for the filament eruption \citep{feynman1995,yan20a}. On the other hand, in the processes of the ideal magnetohydrodynamic instabilities, any further change of the instability-related parameter in the magnetic field, e.g., the magnetic twist, the decay index of the overlying field \citep{che17}, would trigger the eruption. Performing MHD simulations, it had found that a magnetic flux rope becomes kink-unstable when the twist number reaches about 1.8 turns \citep{tor04,fan05}. While the decay index of the background magnetic field reaches about 1.5, the magnetic flux rope will undergo torus instability and erupt successfully \citep{kli06,aul10}.

Numerous efforts have been devoted to unraveling the initiation of solar eruptions. \cite{cheng13,cheng20} proposed that the torus instability serves as the initiator and possibly impels the primary acceleration of the eruption. Some researchers argue that the tether-cutting reconnection mechanism is responsible for the slow-rise phase, while the main acceleration is attributed to torus instability \citep{woods2018,che18,cheng2023}. \cite{jing2021} suggested that both torus and kink instabilities are crucial for the initiation in their study case. Furthermore, \cite{ishiguro2017} introduced a novel type of instability known as a double-arc instability (DAI), wherein a sigmoidal configuration created by a double-arc electric current system destabilizes without weakening the overlying magnetic fields. Some observations are believed to be triggered by this mechanism \citep{kang2019, kim2022}. Recently, \cite{jiang2021} argued that magnetic reconnections driven by photospheric shearing motion are significant in the initiation of solar eruptions. Despite the fact that most observed events can be interpreted with one or more of these models, the onset mechanism of solar eruptions remains a subject of intense controversy.

To gain a deeper understanding of the onset mechanism of eruptions, particularly involving active region filaments, we investigate an eruption of an inverted U-shaped solar filament in Active Region NOAA 12680 in this paper. By using stereoscopic and high-resolution observations, we study the initialization of filament eruption and analyze two overlying precursor activities before the eruption. Furthermore, we also discuss and probe the onset mechanism of the filament eruption. The sections of this paper are organized as follows: observations and data are described in Section 2, main analysis results are presented in Section 3, and the summary and discussions are given in Section 4.

\section{Data analysis and methods} \label{sec:obser-meth}
\subsection{Data analysis}
In this study, the data set is primarily from the New Vacuum Solar Telescope\footnote{\url{http://fso.ynao.ac.cn}} \citep[NVST;][]{liu14,yan20b}, the Solar Dynamics Observatory\footnote{\url{https://sdo.gsfc.nasa.gov}} \citep[SDO;][]{pen12}, the Solar Terrestrial Relations Observatory-Ahead\footnote{\url{https://stereo.gsfc.nasa.gov}} \citep[STEREO-A;][]{kai08}. NVST is a vacuum solar telescope with a 985 mm clear aperture located at Fuxian Lake, in Yunnan Province, China. It can provide high-resolution images of the H$\alpha$ band with a cadence of 11 seconds and the CCD plate scale of  of 0.\arcsec 165 pixel$^{-1}$. The H$\alpha$ images are recorded by a tunable Lyot filter with a bandwidth of 0.25 $\rm\AA$ and the field of view (FOV) is about 150\arcsec $\times$ 150\arcsec. The H$\alpha$ images are reconstructed by using the speckle masking method \citep{xia16}. All NVST H$\alpha$ images are normalized by the quiet Sun and aligned with each other based on a cross-correlation algorithm \citep{yang15}. The Atmospheric Imaging Assembly \citep[AIA;][]{lem12} and the Helioseismic and Magnetic Imager \citep[HMI;][]{sch12,hoe14} on board the SDO can provide full-disk, multiwavelength, high spatio-temporal resolution imaging and magnetic data for this study. The extreme ultraviolet (EUV) images of 304, 171, 193 $\rm \AA$ from SDO/AIA are utilized to show the information from the high chromosphere to the corona. They have a cadence of 12 s and a spatial resolution of 1.\arcsec 2. The line-of-sight (LOS) magnetic fields and Vector magnetograms from the SDO/HMI are employed to show the magnetic information on the photosphere. Their spatial resolutions are 1.\arcsec 0, while their cadences are 24 s and 12 mins, respectively. The AIA images were derotated to a reference time of 07:30 UT to remove the effect of the Sun’s rotation and differential rotation while the NVST H$\alpha$ are carefully co-aligned with SDO/AIA 304 $\rm\AA$ images by matching specific features observed simultaneously. 195 $\rm\AA$ EUV images from the Extreme Ultraviolet Imager \citep[EUVI;][]{wue04} on board the STEREO-A are utilized to show the event in the direction of angle 128$\degr$ with the Sun-Earth direction. Their cadences and spatial resolutions are 2.5 min and 3.\arcsec 2, respectively. 
\subsection{Methods} \label{methods}
To understand the magnetic properties associated with the filament eruption, we reconstruct the coronal magnetic field by a nonlinear force-free field (NLFFF) model based on the observed photospheric magnetic field. NLFFF extrapolation is performed by using the ``weighted optimization'' method \citep{whe00,wie04,wie12} with the vector magnetograms observed by SDO/HMI. Before the extrapolation, a 2 × 2 rebinning of the boundary data to 0.72 Mm pixel$^{-1}$.  In this study, the region for extrapolation is within a box of 318 $\times$ 266 $\times$ 266 uniform grid points, which corresponds to about 231 $\times$ 193 $\times$ 193 Mm$^3$. The region can cover the full active region of interest, which is needed for the NLFFF model \citep{der09}. Although there are some limitations of force-free magnetic field extrapolations by neglecting the effects of the plasma $\beta$ (the ratio of gas pressure to the magnetic pressure) that might against the actual situation of the corona, the NLFFF extrapolation might capture the magnetic structure and connectivity of the coronal magnetic field \citep{pet15}. Based on the three dimensions simulation data of the outer solar atmosphere by using the MHD Bifrost Model \citep{car16}, \citet{fle17,fle19} evaluated the nonlinear force-free field (NLFFF) reconstructions with different methods and different boundary conditions. They found that extrapolations from a force-free chromospheric boundary produce measurably better results than those from a photospheric boundary and any chromospheric magnetic field data can measurably improve the reconstruction of the coronal magnetic field. In recent observations, reliable chromospheric magnetic fields are still hard to capture. In addition, the measurements of the photospheric magnetic field contain inconsistencies and noise, particularly in the transverse components. Therefore, it is acceptable that a preprocessing procedure is used for the SDO/HMI photospheric vector magnetograms to drive the observed non force-free data towards suitable boundary conditions for a force-free extrapolation \citep{wie06}. The input parameters ($\mu_1$, $\mu_2$, $\mu_3$, $\mu_4$) are used in the default set as [1, 1, 0.001, 0.01] for the preprocessing procedure. 
To assess how force free is in the photospheric boundary input, the force ($\epsilon _{force}$) and torque ($\epsilon _{torque}$) balance parameters have been calculated, which are defined by the following equations, respectively \citep{wie06}:
\begin{equation}
   \epsilon _{force}=\frac{|\int_S  B_xB_ydxdy|+|\int_S B_yB_zdxdy|+|\int_S(B_x^2+B_y^2)-B_z^2dxdy|}{\int_S(B_x^2+B_y^2+B_z^2)dxdy},
\end{equation}

and
\begin{equation}
  \epsilon_{torque}=\frac{|\int_Sx((B_x^2+B_y^2)-B_z^2)dxdy|+|\int_Sy((B_x^2+B_y^2)-B_z^2)dxdy|+|\int_SyB_xB_z-xB_yB_zdxdy|}{\int_S\sqrt{x^2+y^2}(B_x^2+B_y^2+B_z^2)dxdy},
\end{equation}

\noindent where $B_x$, $B_y$, and $B_z$ are the three components of the vector magnetic field. $(x, y)$ is the position of the magnetic field. A vector magnetogram is consistent with the force-free assumption if: $\epsilon _{force}\ll 1$, $\epsilon _{torque}\ll 1$. The $\epsilon _{force}$ and $\epsilon _{torque}$ of the photospheric boundary are estimated to be about 0.3-0.4 before preprocessing. After preprocessing, the $\epsilon _{force}$ and $\epsilon _{torque}$ decrease sharply to on the order of $10^{-4}$ and $10^{-3}$, which means that the photospheric boundaries inputted for NLFFF extrapolation after preprocessing are consistent with the force-free condition. 

Otherwise, the current-weighted average of the angles $<\theta_i>$ are less than 10\degr, while the $<|f_i|>$ are less than 4$\times$10$^{-4}$. Thus, the extrapolated magnetic fields are satisfied both the force-free and divergence-free conditions \citep{whe00}. 

Based on the extrapolated magnetic fields of NLFFF models, the twist number ($T_w$) could be calculated by the equation \citep{ber06}:
\begin{equation}
T_w=\int_L\frac{\mu_0J_{||}}{4\pi B}dl=\frac{1}{4\pi}\int_L \alpha dl,
\end{equation}
where $\mu_0$, $J_{||}$, $B$ ,$\alpha$ and $l$ are the magnetic permeability of the vacuum, the electric current component parallel to magnetic field, the strength of the magnetic field, the force-free parameter along the magnetic field line and the length of the magnetic field line.
The codes elaborated in \cite{liu16} are used to calculate $T_w$ in our study.

The magnetic pressure in the corona could be calculated by the following equation:
\begin{equation}
P_m=\frac{B^2}{2 \mu_0},
\end{equation}
where $B$ is the strength of the magnetic field derived by the NLFFF model.

If the constraining field declines with height rapidly enough, the flux rope/filament would become torus unstable \citep{kli06,olm2010,kli14}. Numerous observational studies indicate that torus instability play a significant role in triggering the eruptions \citep[e.g.,][]{cheng13,zuc2014,cheng20,kang2023}. To evaluate the decaying characteristic of the coronal magnetic field around the filament, we also calculate the decay index ($n$), which is defined by the following equation \citep{bat78}:
\begin{equation}
  n=-\frac{d \rm ln{\it B_t}}{d \rm ln{\it h}},
\end{equation}
where $B_t$ is the transverse component of the potential field and $h$ is the height above the photosphere. The potential fields (PF) in the corona are obtained by the Green function method with the vertical component of the photospheric vector magnetograms.

\section{result}
\subsection{The overview of the filament}
\begin{figure*}
\centering
\includegraphics{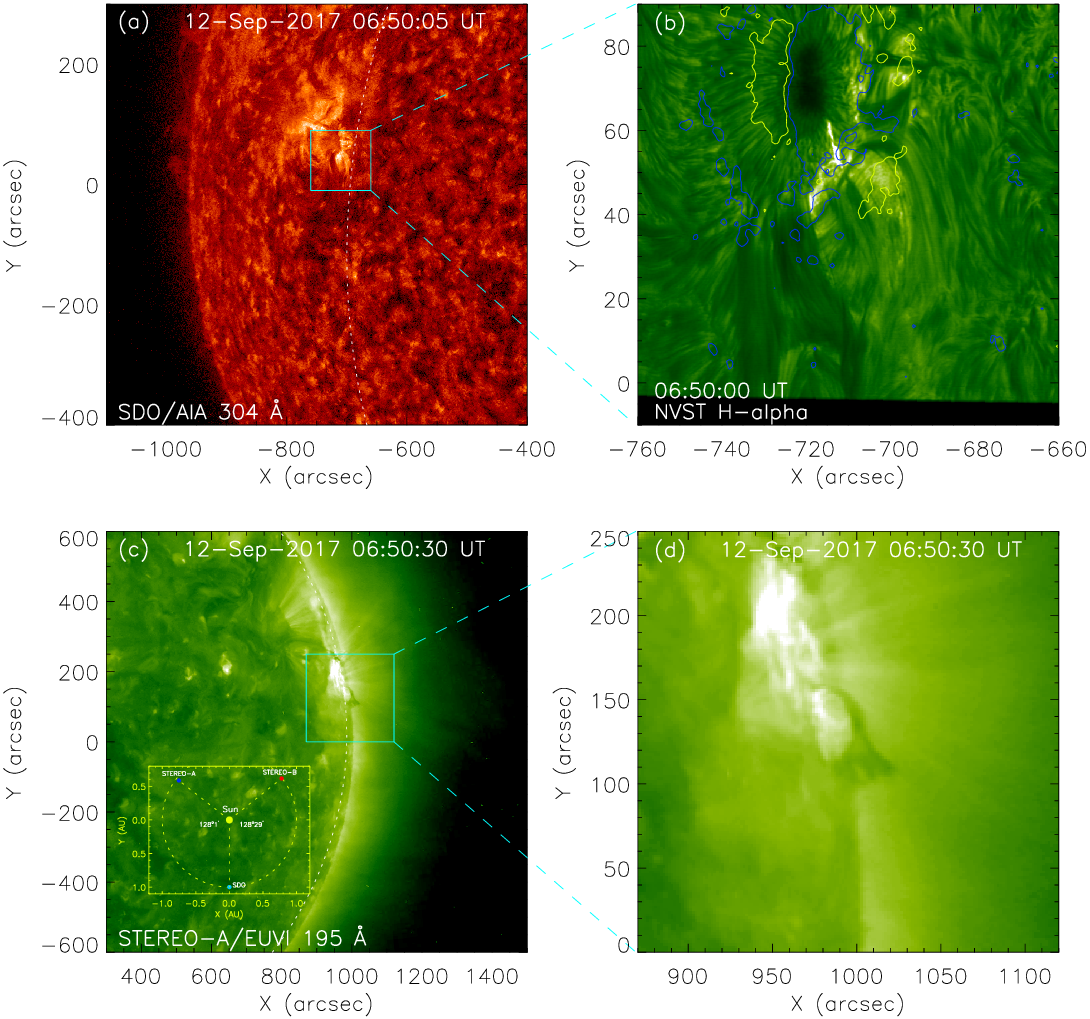}
\caption{Overview observations of the filament. (a): 304 $\rm\AA$ image observed from SDO/AIA. The cyan box outlines the field of view of panel (b). (b): H$\alpha$ images observed from NVST. Yellow and blue contours indicate the positive and negative magnetic field with the levels of $\pm$ 100 G. (c): Observations of STEREO-A/EUVI in 195 $\rm\AA$. The yellow box in the bottom-left corner shows the relative locations of STEREO-A, STEREO-B, and SDO at 06:50 UT. The cyan box outlines the field of view of panels (d). (d): Zoomed-in region of the panel (c). The white dotted lines in panels (a) and (c) denote the same position on Sun. An animation is available, which shows the observation of NVST H$\alpha$ during the period from 06:20 UT to 08:00 UT.}\label{fig1}
\end{figure*}

At active region NOAA 12680, a filament with many overlying filamentary fibrils inhabits nearby a sunspot at around 06:50 UT on September 12, 2017. Fig.\ref{fig1} (a) shows the observation of SDO/AIA 304 $\rm\AA$ while panel (b) is a Zoom-in region of the panel (a) in H$\alpha$ observation obtained by NVST. We can see that the filament has an inverted U-shaped structure and sites at the south of the sunspot with negative polarity. Many filamentary fibrils are crossing over the filament. Fortunately, at this moment, the filament or the active region was sited nearby the solar limb when viewed from the STEREO-A telescope. Fig.\ref{fig1} (c) exhibits the observation of STEREO-A/EUVI 195 $\rm\AA$ band, while panel (d) is a Zoom-in region of panel (c). Viewed by the STEREO-A, it is hard to distinguish the filament on the limb observation. But the arcade-shaped fibrils with many dark materials could be distinguished. Some large-scale coronal loops rooting the sunspot could be also identified (see panel (d)).
\subsection{Two jet-like activities before the filament eruption}
\subsubsection{The processes of two jet-like activities in SDO and NVST observations}
\begin{figure*}[t!]
\centering
\includegraphics{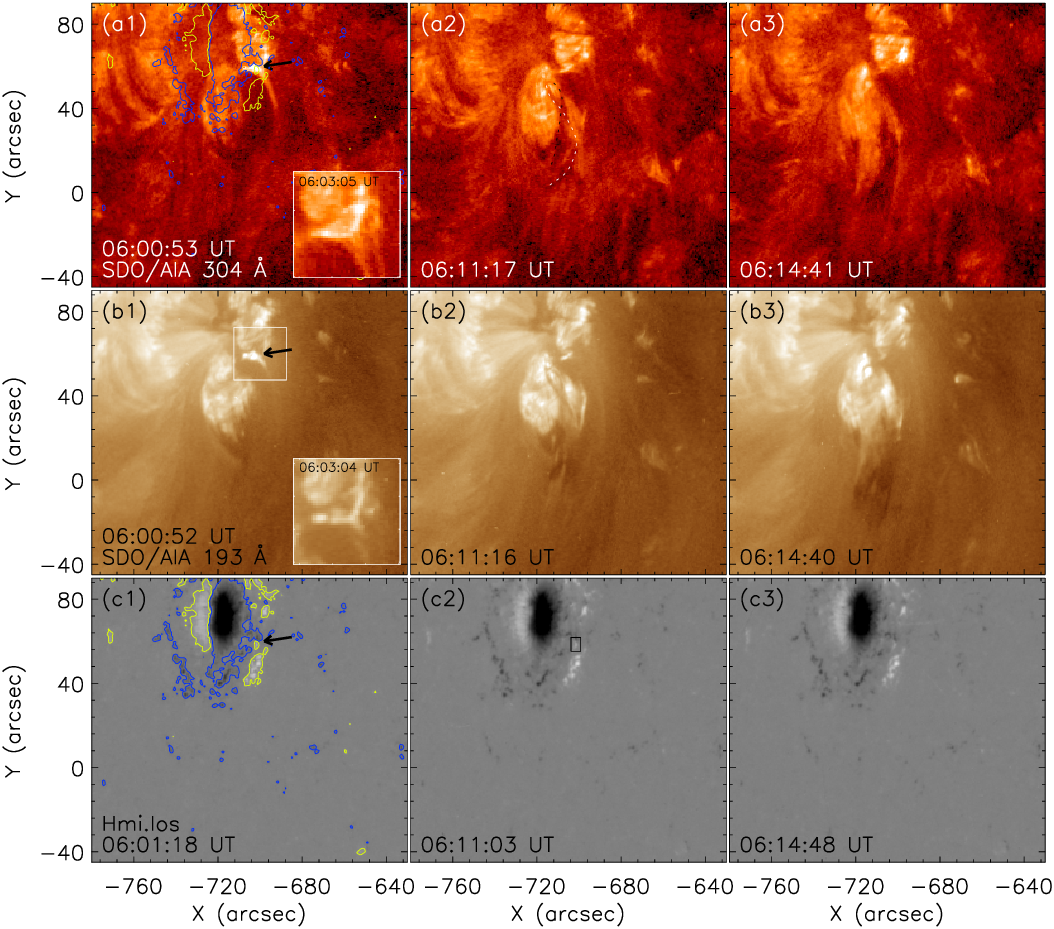}
\caption{The first jet-like activity. (a1)-(a3): 304 $\rm\AA$ images observed by SDO/AIA. The white and black dotted lines outline the twisted structures. (b1)-(b3): 193 $\rm\AA$ images observed by SDO/AIA. The subgraphs in panels (a1) and (b2) are the corresponding images at around 06:03 UT. The white box in panel (b1) outlines the FOV of these two subgraphs. (c1)-(c3): Line-of-sight magnetic magnetograms from SDO/HMI. Yellow and blue contours indicate the positive and negative magnetic fields with the levels of $\pm$ 100 G, respectively. The black box in panel (c2) outlines the region of the magnetic flux in Fig.\ref{fig2} (c). An animation is available, which shows the first jet-like activity in 304 and 193 $\rm\AA$ wavelengths during the period from 05:55 UT to 06:18 UT.}\label{fig2}
\end{figure*}

The fist jet-like activities occurred at around 06:00 UT. Fig.\ref{fig2} shows the detailed process of the first jet-like activity. The upper and middle rows are the observations of SDO/AIA 304 $\rm\AA$ and 193 $\rm\AA$, while the bottom row is the LOS magnetograms from SDO/HMI. Many dark fibrils were lain above the filament at 06:00 UT (see penal (a1)). Some brightenings marked by the black arrows first occurred at the north of the filament (see panels (a1)\&(b1)). At around 06:03 UT, the `Y'-shaped brightening structure could be found in AIA 304 $\rm\AA$ and 193 $\rm\AA$ wavebands (see the sub-region in the panels (a1)\&(b1)). With the occurrence of some brightenings, some overlying fibrils gradually slid along the filament and break away from the initial place. And then, the twist was released along the lifted fibrils as shown by the untwisting motions (see panels (a2)-(a3) \& (b2)-(b3) and the animation of Fig.\ref{fig2}). The white and black dotted lines in panel (a2) outline the winding structures during the release of the untwisted rotations. On the photosphere, some magnetic cancellation took place in the vicinity of the brightening (see panels (c1)-(c3) of Fig.\ref{fig2}).

\begin{figure*}[t!]
\centering
\includegraphics{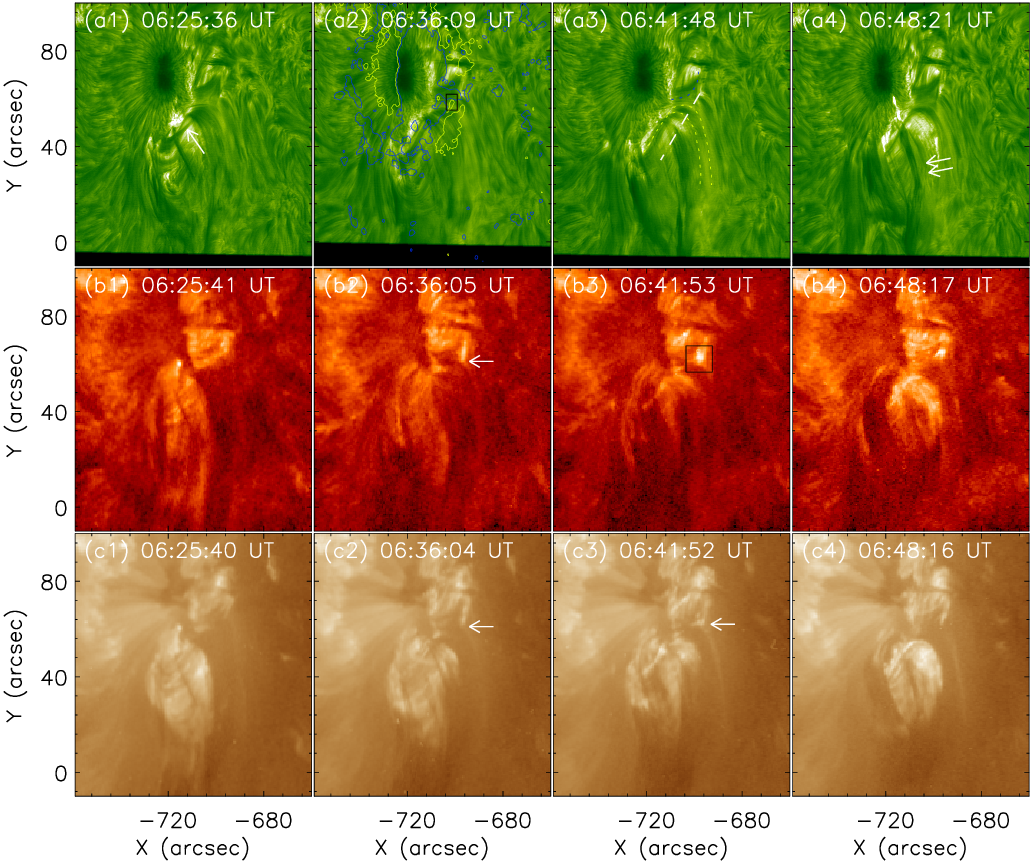}
\caption{The second jet-like activity. (a1)-(a4): H$\alpha$ images observed by NVST. Yellow and blue contours in panel (a1) indicate the positive and negative magnetic fields with the levels of $\pm$ 100 G, respectively. (b1)-(b4): 304 $\rm\AA$ images observed by SDO/AIA. (c1)-(c4): 193 $\rm\AA$ images observed by SDO/AIA. An animation is available, which shows the second jet-like activity in 304 $\rm\AA$ and 193 $\rm\AA$ wavelengths during the period from 06:32 UT to 07:12 UT.}\label{fig3}
\end{figure*}

After half an hour, the second jet-like activity took place at the same place at around 06:40 UT. Fig.\ref{fig3} shows the detailed evolution of the second jet-like activity. Panels (a1)-(a4) are the H$\alpha$ observations obtained from NVST, while panels (b1)-(b4) and (b1)-(b2) are the corresponding observations of SDO/AIA 304 $\rm\AA$ and 193 $\rm\AA$, respectively.
Around 06:25 UT, numerous inverted U-shaped fibrils are identifiable in the interesting region (see panel (a1)-(a2), (b1)-(b2)\&(c1)-(c2)). Notably, there are several bunches of dark fibrils situated above the inverted U-shaped fibrils, marked by the arrow in panel (a1). With the contours of magnetic field displayed in panel (a2), we consider that these dark fibrils were rooted in the positive polarities in the black box of panel (a2). In the context of frozen-in plasma circumstances, these dark fibrils may represent magnetic fields, constraining the inverted U-shaped fibrils.

At first, some brightenings/``Y-shaped'' brightening structures could be distinguished at some place as the first jet-like activities (see the white arrows in panels (b2), (c2)\&(d2)). The constrained dark fibrils disappeared as the brightenings occurred, a phenomenon we attribute to the heating of cool plasma induced by magnetic reconnection. Additionally, many new moving fibrils formed around the brightenings, marked by the yellow and blue dotted lines in panel (a3) (also visible in the animation of Fig.\ref{fig1}). Subsequently, these newly formed moving fibrils, accompanied by the inverted U-shaped fibrils containing dark materials, ascended generally with some distinct untwisted rotations (see panels (a3)-(a4) \& (b3)-(b4) and the animation of Fig.\ref{fig3}). The lifting fibrils are indicated by white arrows in panel (a4). Eventually, a inverted U-shaped filament had been presented its distinct appearance as these overlying fibrils were released (see panel (a4)). 

\begin{figure*}[t!]
\centering
\includegraphics{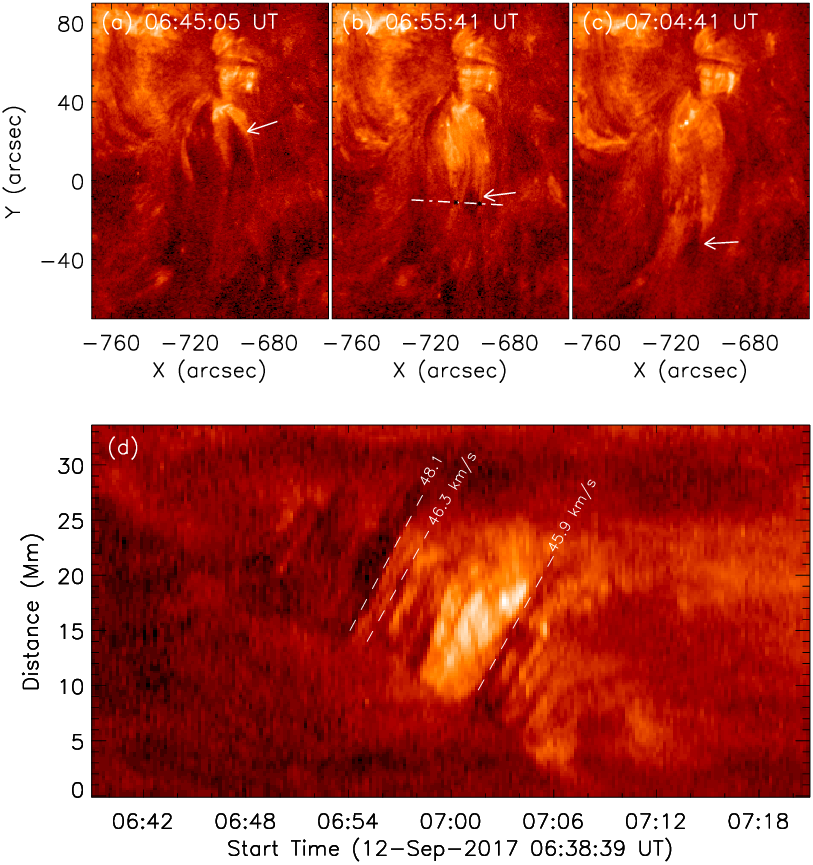}
\caption{Estimation of the twist number released by the second jet-like activity. (a)-(c): SDO/AIA 304 $\rm\AA$ images at different moment. The white arrows mark the dark structure at a different moment. The white dotted-dashed line outlines the path for the time-distance diagram of panel (d). Two black asterisks on the white dotted-dashed line mark the width of the rotation. (d): Time-distance diagram reconstructed by SDO/AIA 304 $\rm\AA$ observations along the white dotted-dashed line in panel (b). Dashed lines mark some dark inclined stripes.}\label{fig4}
\end{figure*}

To quantify the untwist number released by the second activity, we trace the special structure to estimate the twist number from the untwisted rotations. Panels (a)-(c) of Fig.\ref{fig4} show the untwisted rotation revealed by SDO/AIA 304 $\rm\AA$ observations. By tracking the dark structure marked by the white arrows, we derive that the dark structure had rotated at least two rounds (also see the animation of Fig.\ref{fig3}). This means that the untwist number released by the second activity can be more than two turns. On the other hand, we also make a time-distance diagram along the white dotted-dashed line perpendicular to the rolling tube. Panel (d) exhibits the time-distance diagram derived from a series of SDO/AIA 304 $\rm\AA$ images. There are many distinct inclined stripes, which means that many dark structures went along the dotted-dashed line during the untwisted rotation. According to the imprint of the inclined stripes, the speed of these structures along the path was estimated to be about 47 km/s. With the assumption that untwisted rotation kept a constant speed and in a constant radius, The untwist number released by the activity could be estimated by the formula of $T_n = vt/2\pi r$, where $v$, $t$ and $r$ are the speed, duration time, and the radius of the rotation, respectively. The speed ($v$) of the rotation can be considered to be 47 km/s, derived from the time-distance diagrams. The activity was experienced during the period from around 06:45 UT to 07:06 UT, so $t$ equals about 21 minutes. The width ($d$) of the rotation marked by two dark asterisks in panel (b) can be estimated as about 8.7 Mm, which indicates that $r = d/2$ is about 4.35 Mm. According to the above values of parameters and the formula, we can obtain that $T_n$ is about 2.1 turns which is consistent with the estimation based on tracing the special structure. 

\subsubsection{Some main observational phenomenons and interpretations}
\begin{figure*}[t!]
\centering
\includegraphics{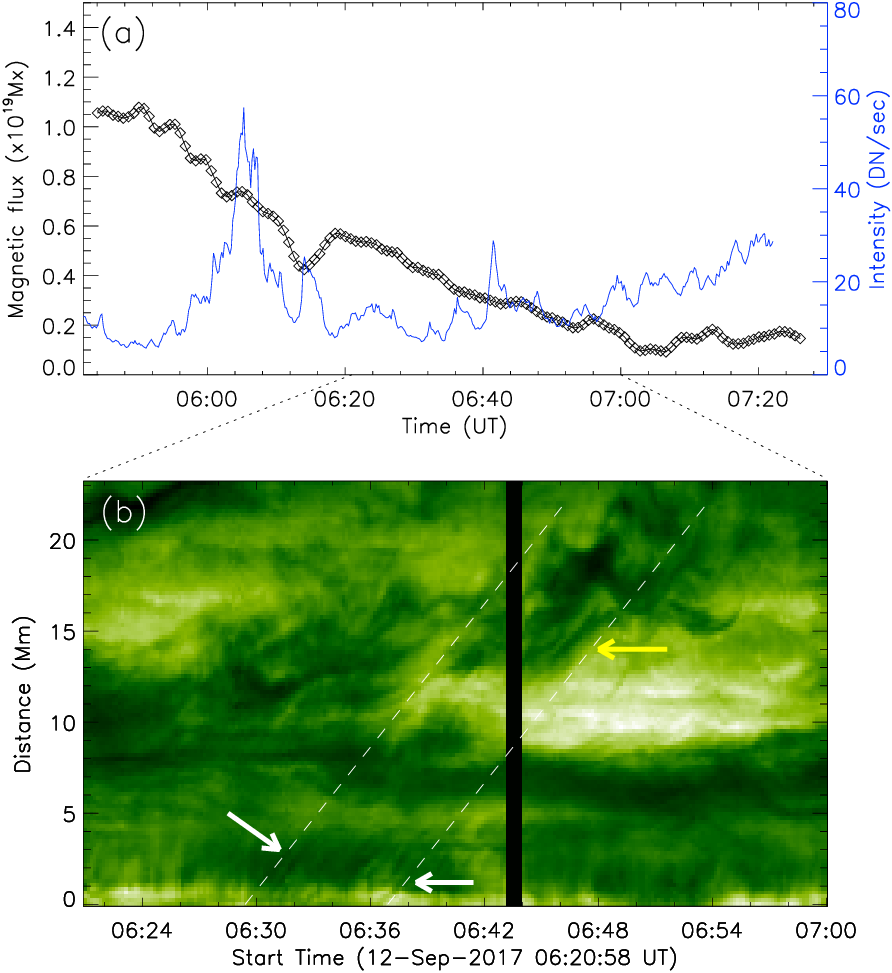}
\caption{Variations of some parameters. (a): Variations of positive magnetic flux in the black box of Fig.\ref{fig1} (c2) and Fig.\ref{fig2} (a1) and SDO/AIA 304 $\rm\AA$ mean intensities in the box of Fig.\ref{fig2} (b2). The mean intensities are normalized by the duration. (b): Time-distance diagram derived from a series of NVST H$\alpha$ images along the dashed line in Fig.\ref{fig2} (a2).}\label{fig5}
\end{figure*}

\begin{figure*}[t!]
\centering
\includegraphics{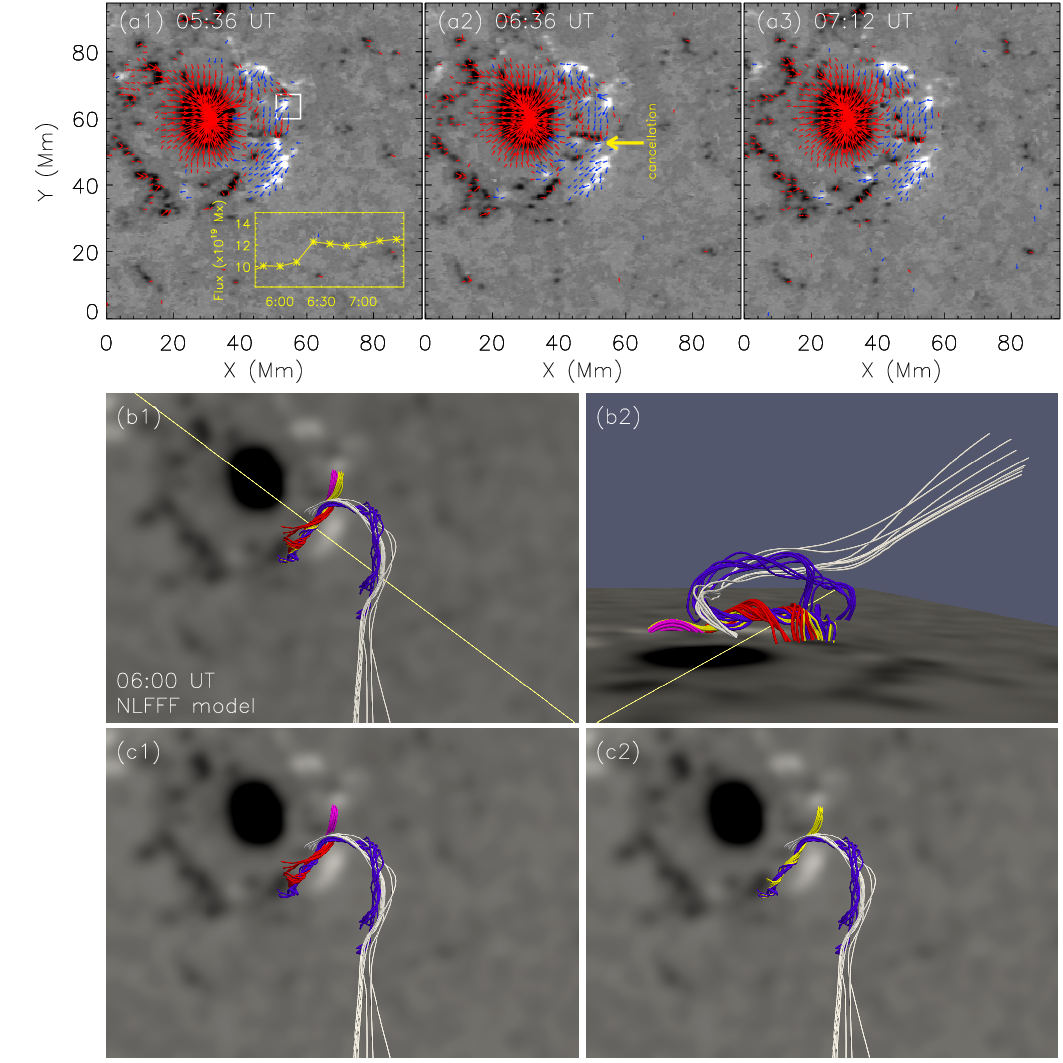}
\caption{Magnetic structures associated with two jet-like activities. (a1)-(a3): Vector magnetograms. The blue and red arrows denote the transverse magnetic field with positive and negative vertical magnetic fields, respectively. The yellow box in panel (a1) shows the variation of the positive magnetic flux in the white box of panel (a). (b1)-(b2) and (c1)-(c2): Magnetic field lines derived by the NLFFF extrapolation at 06:00 UT. Panels (b1), (c1) \& (c2) are a top view, while panel (b2) is a side view. An animation is available, which shows the evolution of vector magnetograms during the period from 04:00 UT to 07:48 UT.}\label{fig6}
\end{figure*}

Fig.\ref{fig5} (a) shows the variations of positive magnetic flux in the black box of Figs.\ref{fig2} (c2) \& \ref{fig3} (a2) and the mean intensity of AIA 304 $\rm\AA$ in the box of Figs.\ref{fig3} (b3). The mean intensity is normalized by its duration. The positive magnetic flux exhibited a decreasing trend, which evidences the magnetic cancellation between the positive flux and upper negative flux (see Fig.\ref{fig2} (c1)-(c3) and the animation of Fig.\ref{fig6}). Otherwise, there are two peaks of AIA 304 $\rm\AA$ intensity at around 06:05 UT and 06:41 UT, which indicates the brightenings caused by two jet-like activities. Panel (b) shows the time-distance diagram reconstructed by the NVST H$\alpha$ images along the white dashed line in panel (a2) of Fig.\ref{fig3}. We could see that many inclined structures marked by the two white arrows appeared near the bottom of the time-distance diagram. Some dark structures marked by the yellow arrow at around 06:48 UT are the signal of material fibrils associated with the second jet-like activity. According to their inclination outlined by two white dashed lines, these inclined structures and dark structures are closely related phenomena. Also evidenced by the animation of Fig.\ref{fig1}, it could manifest that the second jet-like activity originated from the brightenings marked by Fig.\ref{fig3} (b2), (c2)\&(c3) and was associated with the newly formed and moving fibrils.

In Fig.6 (a1)-(a3), vector magnetograms from SDO/HMI are presented. It is observed that some magnetic flux emerged around the cancellation site. The yellow line depicts the variation of positive flux within the white box of Fig.6 (a1). These positive emerging fluxes were associated with negative fluxes located to the north and in proximity to the positive flux of cancellation (see the direction of the magnetic field, indicated by the blue/red arrows). Therefore, we consider that it was newly emergence of dipoles. The negative flux from the emergence had undergone cancellation with the positive flux rooting the constrained magnetic field, as marked by the yellow arrow in Fig.\ref{fig6} (a2).
By 07:12 UT, it is evident that a substantial portion of the positive flux, indicated by the yellow arrows in panel (a2), had been canceled almost entirely. That is consistent with the result of Fig.\ref{fig5} (a). 
Fig.\ref{fig6} (b1)-(b2) shows some selected magnetic field lines derived by the NLFFF model at 06:00 UT. Panel (b1) shows the magnetic field lines from top view while panel (b2) is the side view. The blue lines denote the magnetic field of the filament, while the white lines indicate some open magnetic fields sited near the filament. The pink lines denote the emerging magnetic field, while the red lines denote the magnetic field rooted in the positive fluxes. The red magnetic field lines are the constraint magnetic fields. 

Based on the fact of some newly emergence of dipoles nearby the positive polarity and cancellation between the positive polarity and the negative polarity of the emerged dipole, we suggest the long magnetic field lines (the yellow lines) should be formed by either the elevation of serpentine magnetic field or magnetic reconnection of two bunches of or a ``U-shaped'' magnetic field, resulting in some magnetic cancellation at the junction (see panels (c1)\&(c2)). As the yellow lines reconnected with the open lines (the white lines), it would form two bundles of magnetic field lines (long inverted U-shaped open and short lines, liking the yellow and blue lines in Fig.\ref{fig3} (a2)) and cause the brightenings/ ``Y-shaped'' brightenings at around the reconnection site. The twist in the red/yellow lines transfer to the open lines, which could cause the twist motion to release the twist in open lines likes the event studied by \cite{yang19}. Moreover, based on high-resolution observations, it is evident that some inverted U-shaped fibrils containing dark materials have been also released during the jet-like activity. This suggests that the inverted U-shaped open magnetic structures, accompanied by newly forming inverted U-shaped open lines, were released during the twist motion.

In these scenarios, our observations, including newly emerging dipoles, magnetic cancellations,``Y-shaped'' brightenings, newly forming fibrils, and rotational motion, can be well explained.
Therefore, we infer that these two jet-like activities were caused by the aforementioned scenarios, resulting from magnetic reconnections caused by emerging flux. This process leads to the removal of constraint fields and the release of inverted U-shaped fibrils containing dark materials.

\subsubsection{Two overlying jet-like activities observed from STEREO-A view}
\begin{figure*}[t!]
\centering
\includegraphics{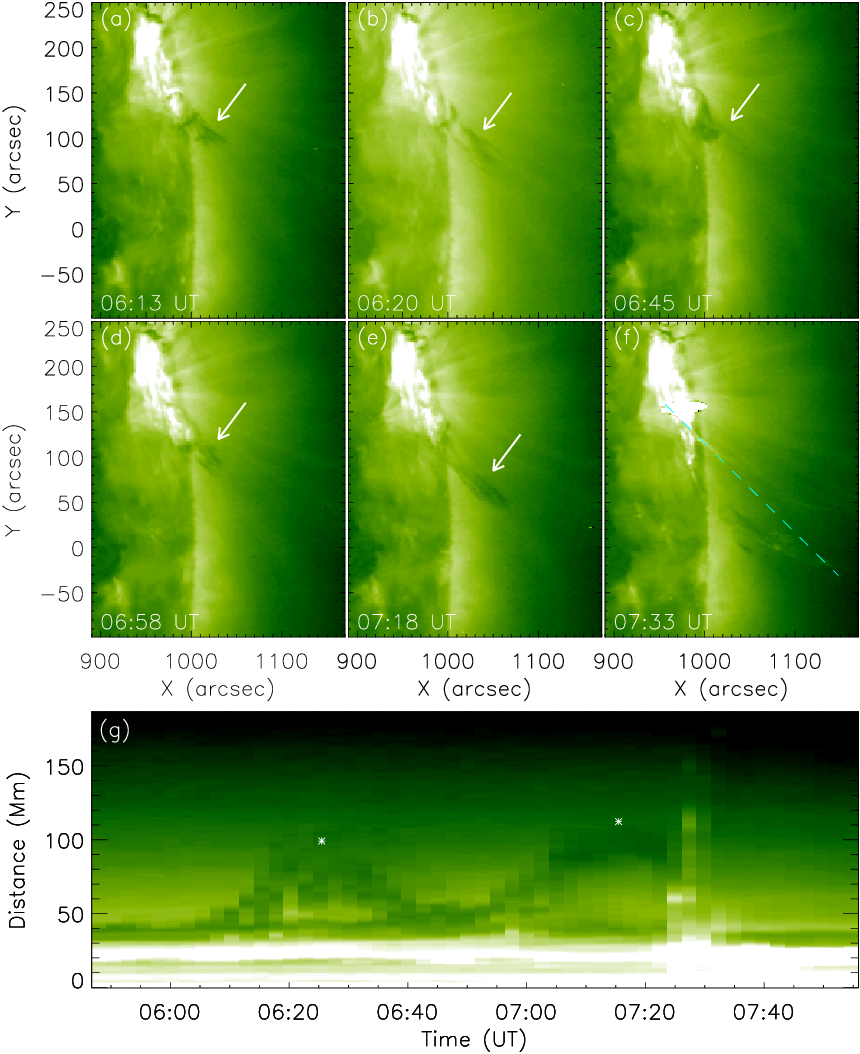}
\caption{STEREO-A 195 $\rm\AA$ observations. (a)-(c): 195 $\rm\AA$ observations for the first jet-like activity. (d)-(f): 195 $\rm\AA$ observations for the second jet-like activity. The white arrows point to the dense dark material lifted by the jet-like activities. (g): Time-distance diagram constructed by the STEREO-A observations along the cyan dashed line in panel (f). Two white asterisks mark the highest points of the lifting dark material.}\label{fig7}
\end{figure*}

At the limb viewed by the STEREO-A observations, many dense materials were ejected behind the dark arcade-shaped fibrils by both two jet-like activities. Panels (a)-(c) of Fig.\ref{fig7} exhibit the 195 $\rm\AA$ observations during the first activity. We can see that many dark materials marked by the white arrows were ejected aslant and fell back along the same path. Panels (d)-(f) of Fig.\ref{fig7} are the 195 $\rm\AA$ observations during the second activity. We can also find the same phenomenon as the first activity. At around 07:33 UT, both dark materials and arcade-shaped fibrils had been thrown away during the eruption  (see panel (f)).

Fig.\ref{fig7} (g) exhibits the time-distance diagram derived from 195 $\rm\AA$ observations along the path of the dashed line in panel (f). The dark materials ejected by the activities showed a two-peak oscillation pattern before the eruption. The eruption occurred when the material fell back incompletely. The height of the two peaks along the slit path could be estimated to be about 99.1$\pm$2.2 Mm and 112.3$\pm$2.2 Mm, which correspond to the heights of the jetted plasma above the solar limb of 43.2$\pm$1.6 Mm and 52.1$\pm$1.6 Mm, respectively. The uncertainties in the distances are estimated as two pixels of the image. Assuming that these materials were only affected by solar gravity after acceleration and experienced a free-fall motion, it requires that the initial speeds of these materials were at least about 149.2$\pm$2.6 and 162.9$\pm$2.4 km/s, respectively. These values are comparable with the speeds of the plasma along the magnetic field during the jets in previous studies \citep{wan18}.

\begin{figure*}[h!]
\centering
\includegraphics{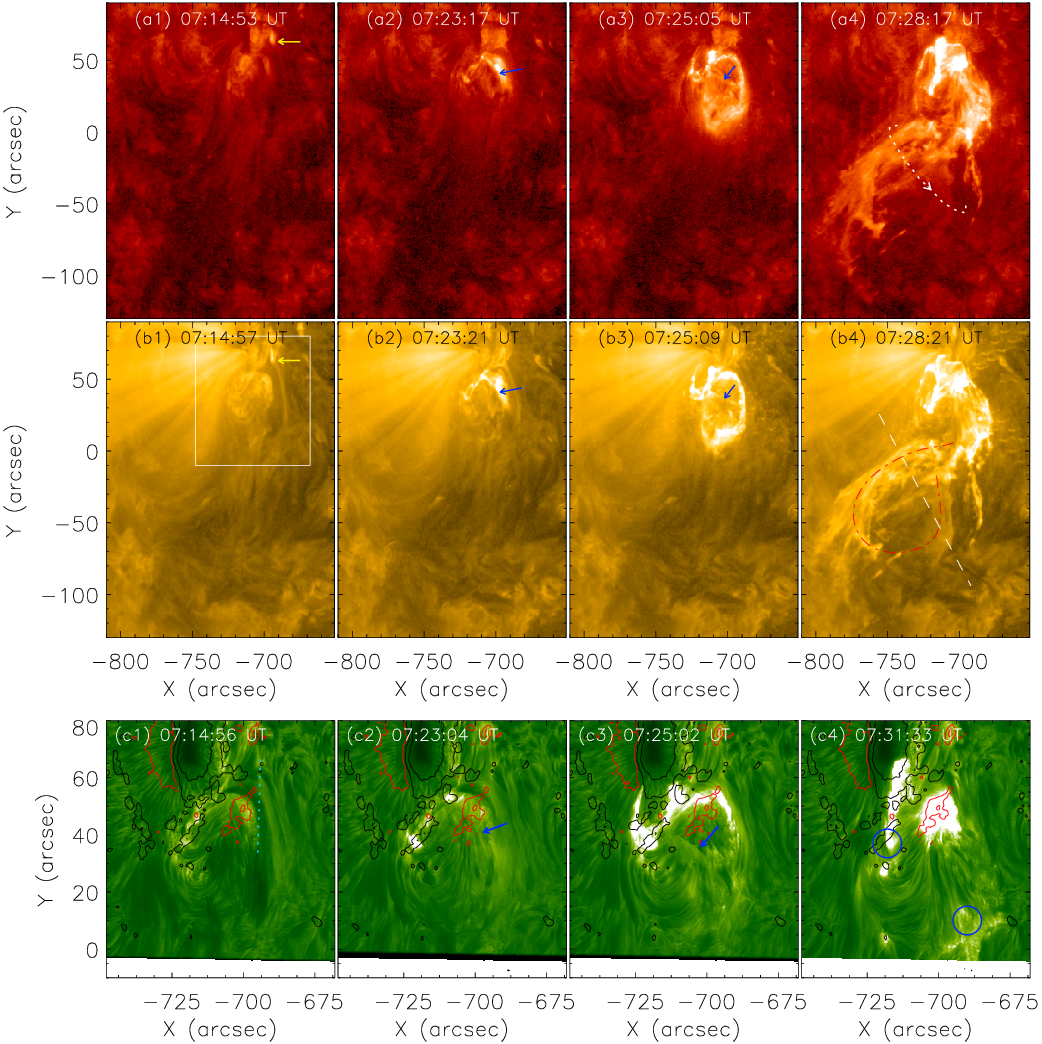}
\caption{The process of the filament eruption. (a1)-(a4): SDO/AIA 304 $\rm\AA$ observations. (b1)-(b4): SDO/AIA 171 observations. The white box outlines the field of view of panels (c1)-(c4). (c1)-(c4): NVST H$\alpha$ observations. The blue arrows point out the erupting filament. Red and black contours indicate the positive and negative magnetic field with the levels of $\pm$ 100 G. Blue circles mark the brightenings associated with the footpoints of the inverted U-shaped filament. An animation is available, which shows the eruption process of the filament in 304 $\rm\AA$, and 171 $\rm\AA$ wavelengths during the period from 07:12 UT to 07:52 UT.}\label{fig8}
\end{figure*}

\subsection{The eruption process of the filament} 
\begin{figure*}[t!]
\centering
\includegraphics{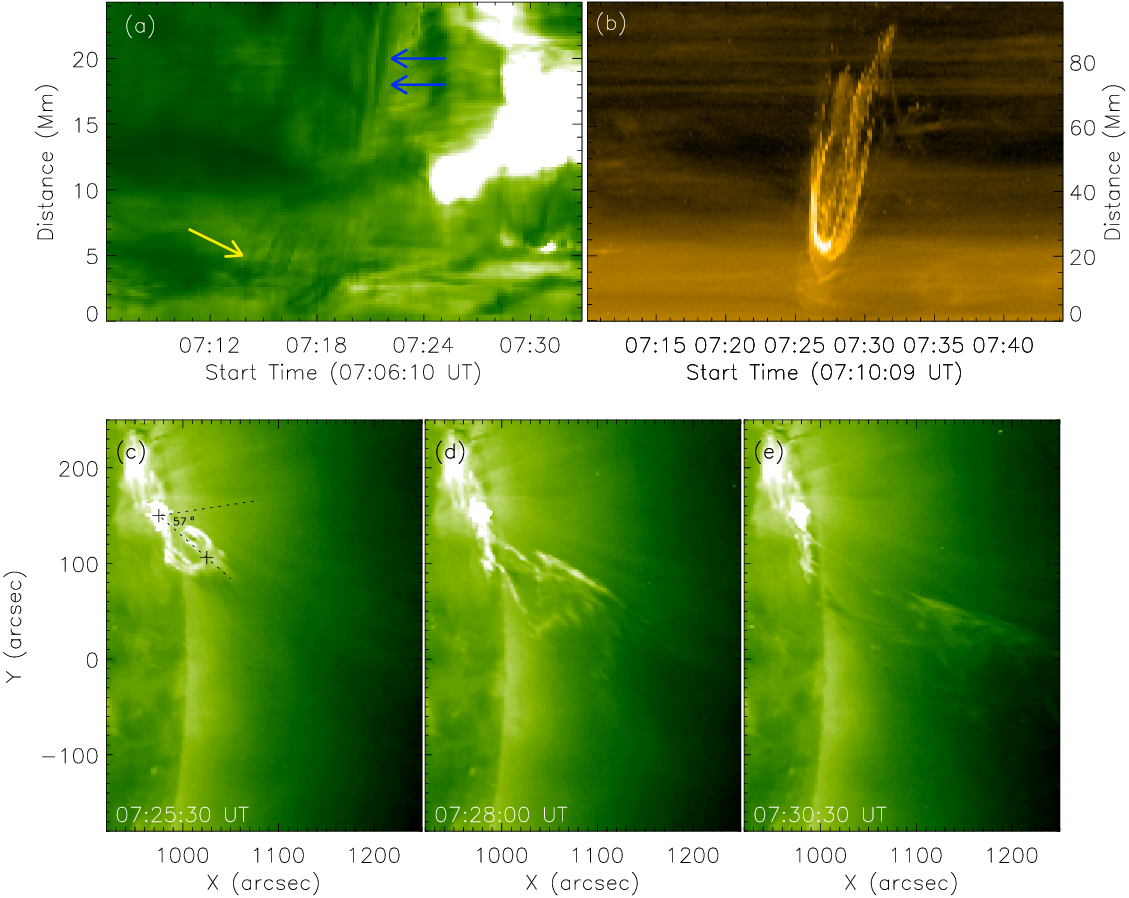}
\caption{The process of the filament eruption with STEREO-A observations. (a): Time-distance diagrams constructed by the H$\alpha$ observations alone the light blue dotted solid lines in panels (c1) of Fig.\ref{fig8}. (b): Time-distance diagrams constructed by the SDO/AIA 171 $\rm\AA$ observations alone the white dashed line in panels (b4) of Fig.\ref{fig8}. (c)-(e): STEREO-A 195 $\rm\AA$ observations.}\label{fig9}
\end{figure*}

After the second jet-like activity, the filament became unstable and commenced its eruption around 07:20 UT. Fig.\ref{fig8} illustrates the eruption process of the filament. Initially, some brightenings occurred in the cancellation site as the same of the two jet-like activities around 07:14 UT, as indicated by the white arrows in panels  (a1)\&(b1). Intermittently, there were newly moving flows ejected from the brightening point (see panel (c1) and the animation of Fig.\ref{fig8}). Panel (a) of Fig.\ref{fig9} shows the time-distance diagram along the blue dotted line in panel (c1) of Fig.\ref{fig8}. It is noticeable that numerous dark stripes, marked by the yellow arrow, appeared between 07:14 UT and 07:18 UT before the filament started lifting, as indicated by the two blue arrows. This observation provides evidence that newly moving flows were intermittently ejected before the filament lifting. Combining the brightenings and these moving flows, we deduce that it is the same physical process involving magnetic reconnection as the two jet-like activities before the eruption, which also leads to a decrease in the constrained field.

Following that, the inverted U-shaped filament lifted from its middle part, as indicated by the blue arrows (see panels (a2)-(a3), (b2)-(b3), and (c2)-(c3)), with some brightenings occurring around the filament. Subsequently, it ejected away with an apparent rotational motion, its main axis marked by the red dotted-dashed line in panel (b4). The rotation motion was clockwise when viewed from the top (see panels (a4), (b4), and the animation of Fig.\ref{fig8}). Panel (b) of Fig.\ref{fig9} presents the time-distance diagram constructed from SDO/AIA 171 $\rm\AA$ images along the dashed line in Fig.\ref{fig8} (b4). Many bright stripes in the diagram also indicate the rotation motion during the eruption.
In previous studies, this type of rotational motion is typically attributed to the eruption of a highly twisted flux rope \citep{amari2000, fan2007}.
Furthermore, it could be inferred that the progenitor of the eruption, the filament, should have a highly twisted structure. At around 07:31 UT, two footpoints of the filament had also been brightened (marked by two blue circles in panel (c4) of Fig.\ref{fig8}) which are in the south of the flare ribbon. 

Fig.\ref{fig9} (c)-(e) exhibits the eruption process observed by the STEREO-A view. At around 07:25 UT, a tortuous structure representing the filament had been lifted away from the solar surface. After that, this structure had been ejected away along a non-radial path (see panels (d)-(e)). The angle between eruption path and radial direction was estimated to be about 57 degree (see panel (c)). This means that it experienced a non-radial eruption. Two threads connecting the ejecting structure with the solar surface were identified. This also further demonstrates that two footpoints of the filament are in the south of the post-flare loops, which is consistent with the finding of Fig.\ref{fig8} and the extrapolated magnetic structures of the filament (see Fig.\ref{fig5}).

\subsection{The decay index, magnetic pressure, and twist number}
\begin{figure*}[t!]
\centering
\includegraphics{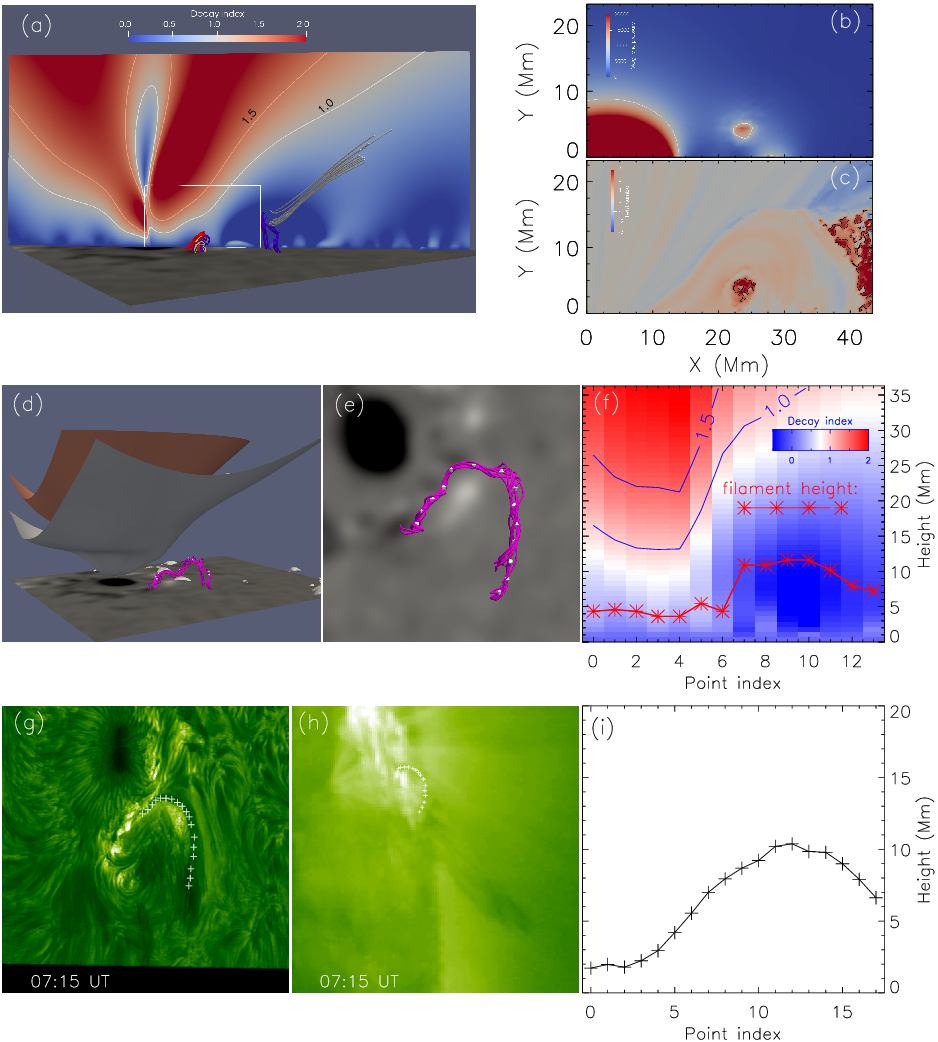}
\caption{The magnetic properties before the filament eruption. (a): The distribution of the decay index in the cross-section, superimposing the selected magnetic field lines from Fig.\ref{fig6}. The cross-section is perpendicular to the solar surface along the yellow line in Fig.\ref{fig6} (b1)\&(b2). The contours indicate the decay index values of 1.5 and 1.0. The white rectangle box outlines the region of panels (b)\&(c). (b): The distribution of magnetic pressure, with white contours denoting magnetic pressure at 10000 dyn/cm$^{-2}$ ($\sim$ 500 Guass). (c): Twist number, with a black contour outlining the twist number of two turns. (d): Selected magnetic field lines representing the inverted U-shaped filament and three-dimensional contours of decay index at 1.5 and 1.0. (e): Selected magnetic field lines representing the inverted U-shaped filament seen from a top view, overlapping with some selected white dots. The white dots in panels (d)\&(e) mark the filament structure. (f): The height of the filament marked by the white dots. The background shows the decay index with height at the corresponding points. The contours represent decay index values of 1.5 and 1.0. (g): NVST H$\alpha$ observaion at 07:15 UT. (h): The corresponding STEREO-A 195 $\rm\AA$ observaion. The white plus signs marks the same structure representing the filament seen from different views. (i): The height of the selected white plus signs.}\label{fig10}
\end{figure*}

According to the methods described in the section \ref{methods}, we calculate the decay index, magnetic pressure, and twist number. Figs.\ref{fig10} (a)-(c) exhibit the distributions of different magnetic parameters at 07:00 UT in a cross-section cutting the filament and perpendicular to the solar surface. The yellow lines in Fig.\ref{fig5} (b1)\&(b2) marks the intersection between the cross-section and the solar surface. Panel (a) shows the decay index in the cross-section and the selected magnetic field lines as Fig.\ref{fig5}, while panels (c)\&(d) are magnetic pressure and the twist number in the white rectangle box of panel (a). It is evident that the filament exhibited a strong magnetic pressure and a high twist number. We consider that the filament was situated in a region with magnetic pressure exceeding 10$^4$ dyn/cm$^2$. Specifically, we derive that the mean magnetic strength (magnetic pressure) of the filament was approximately 540 Gauss (1.16$\times$10$^4$ dyn/cm$^2$), and the mean twist number was around 2.4 turns. This is comparable to the untwist number (2.1 turns) released by the second jet-like activity. 

On the other hand, Fig.\ref{fig10} (a) shows the distribution of the decay index in the cross section, while the three-dimensional appearance is showed in panel (d). At this moment, the entire filament was positioned below the level of 1.5, even below 1.0. For a more in-depth analysis, the selected magnetic field lines representing the filament were marked by 14 white points (see panels (d)\&(e)), which also indicate the location of the filament. Panel (f) displays the height of these white points and overlays the distribution of decay index with height. It is derived that the maximum height of the filament is about 12 Mm. Furthermore, it is evident that the filament was situated in the region below the decay index of 1.0. To ensure accuracy in estimating the filament's height and to avoid inaccuracies in the NLFFF model \citep{fle17}, observations from two perspectives were utilized to estimate the height of the filament around 07:15 UT (see panels (g) \& (h)). By using the routine ``scc\_measure.pro'' in the SSW package \citep{thompson2009, cheng20, guo2021}, the height of the filament was derived by identifying the filament structures in two different viewing observations. The observational filament is marked by the white plus signs in panels (g)\&(h). Panel (i) displays the derived height of the filament, demonstrating that the filament was located below 12 Mm. That is consistent with the height derived by the NLFFF model. Therefore, the torus instability plays a minor role in triggering this filament eruption.

However, as the filament upraise after the filament eruption is initiated, it becomes susceptible to ascending into regions with a decay index exceeding 1.5. Throughout the filament eruption, the reconnection in the flare current sheet below the filament may play a more crucial role in accelerating the eruption, possibly coupled with torus instability, as discussed in previous works \citep{cheng20,cheng2023}. This interplay of processes could explain why the filament eruption was eruptive and not confined.

\begin{figure*}[t!]
\centering
\includegraphics{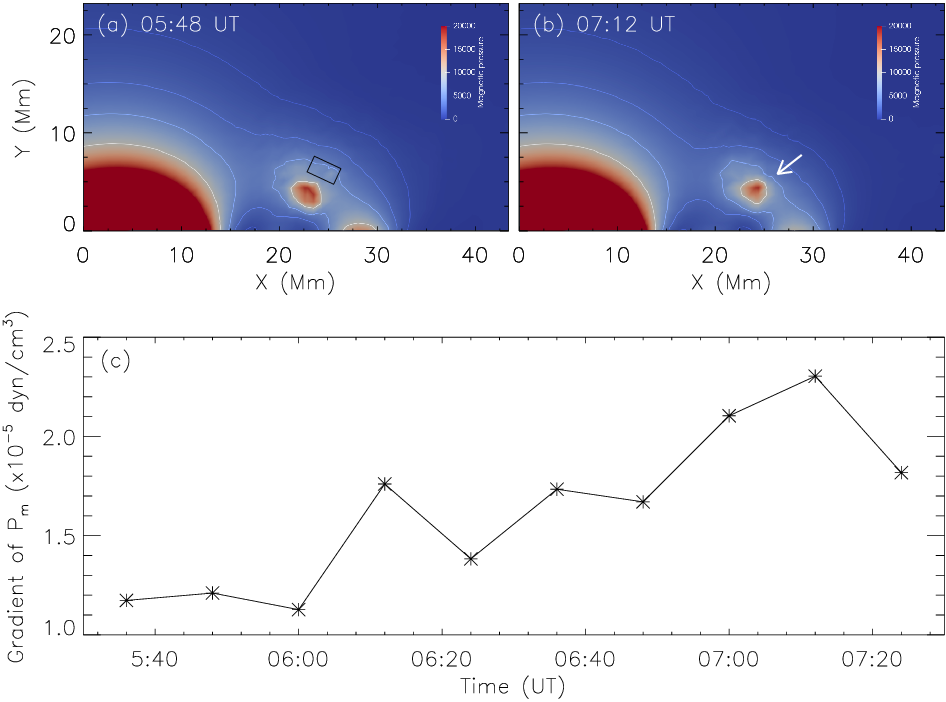}
\caption{The distribution of the magnetic pressure in the white box of Fig.\ref{fig9} (a). (a)-(b): The magnetic pressure at 05:48 UT and 07:12 UT. The contours outline the magnetic pressure with the levels of 1000, 2000, 5000, and 10000 dyn/cm$^2$. (c) The variation of the mean gradient of magnetic pressure in the black box of the panel (a).}\label{fig11}
\end{figure*}

Figs.\ref{fig11} (a) \& (b) show the distribution of magnetic pressure in the Zoom-in region of Fig.\ref{fig10} (a) at 05:48 UT and 07:12 UT, respectively. Compared with these two panels, an attractive shrinkage marked by the white arrow was found over the filament, which means the increase of the gradient of magnetic pressure in that place. To qualitatively analyze, we calculate the mean gradient of magnetic pressure in the black box of Fig.\ref{fig11} (a). Panel (c) exhibits the variation of mean gradient magnetic pressure in the black box. At around 05:48 UT before the two overlying jet-like activities, the mean gradient of magnetic pressure was about 1.2$\times$10$^{-5}$ dyn/cm$^3$, while it is about 2.3$\times$10$^{-5}$ dyn/cm$^3$ at around 07:12 UT after the two overlying jet-like activities. As a whole, the gradient of magnetic pressure displayed a gradual increase with time during the periods of two overlying jet-like activities. This suggests that two overlying jet-like activities releasing the constraining magnetic fields (the red lines in Fig.\ref{fig5}) would increase the gradient of magnetic pressure over the filament. With the increase of the gradient of magnetic pressure over the filament and the decrease of the constraining magnetic fields, it is hard to trap the high-twisted filament to erupt. Therefore, the decrease of the constraint magnetic fields caused by the two overlying jet-like activities is the key reason for the filament eruption.

\section{Summary and discussion}\label{sec:conclusion}
In this paper, we investigate an inverted-U shaped filament eruption and its two precursory activities in active region NOAA 12680 on September 12, 2017. By using different ground-based and space-based telescopes, we study some precursory activities before filament eruption and the process of the its eruption. Based on the magnetograms observed by SDO/HMI and the extrapolations, we also analyzed the magnetic properties to explore the physical mechanism for initiation of the eruption. Our main results are as follows:

(1) Two precursory jet-like activities took place before the eruption of the inverted-U shaped filament, which lifted some dense materials to about 50 Mm above the solar limb. They were caused by the magnetic reconnection caused by the emerging flux. The untwisted rotation was found during both two events, and the twist number of at least 2 turns was released in the second event. 

(2) Before the filament eruption, there are some brightenings and newly moving flows occurring in the same location as the onset of two jet-like activities, indicating the same physical process as two jet-like activities. Based on the observations from the STEREO-A view, the eruption appears to be non-radial.

(3) Based on the extrapolation, it is determined that the filament lay under the height corresponding to a decay index of 1.0, exhibiting strong magnetic fields and a high twist number before its eruption. The gradient of magnetic pressure displayed a gradual increase with time before the filament eruption.

In our study, we observed two jet-like precursors, brightenings, and newly moving flows before the eruption of the filament. These phenomena are attributed to the same physical process involving magnetic reconnection, leading to magnetic cancellation and an increase in the gradient of magnetic pressure (see Figs.\ref{fig5}(a) and \ref{fig11} (c)). All these precursors contribute to the release of constraint magnetic fields situated on the positive polarity within the black box in Fig.\ref{fig2}. Consequently, we consider these events as homologous physical processes and believe they play a crucial role in the onset of the filament eruption.

\begin{figure*}[t!]
\centering
\includegraphics{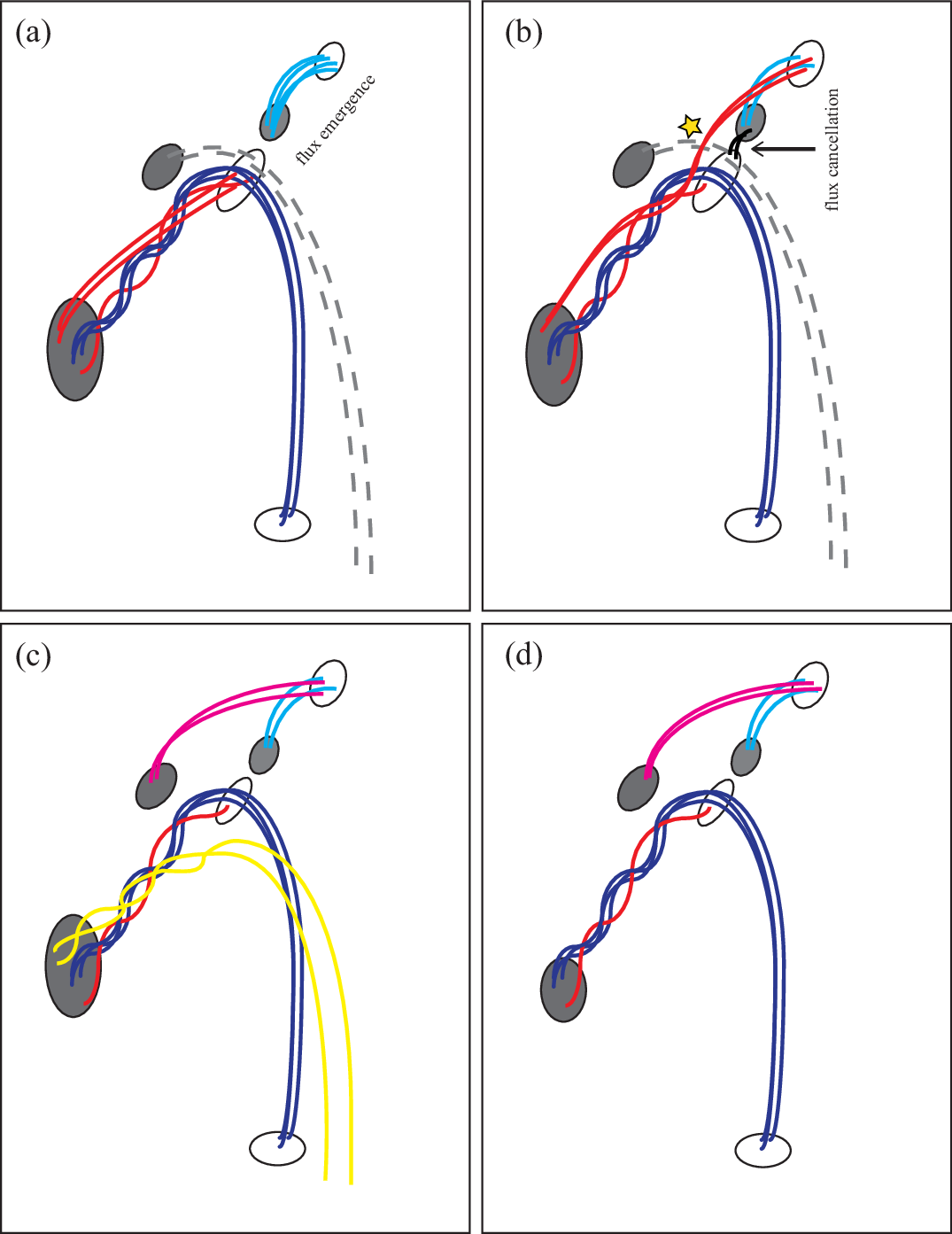}
\caption{Cartoon showing the release of the constraining magnetic field through continuous magnetic reconnection. The cycle filled with gray denote the negative magnetic flux while the cylce filled with white denote positive magnetic flux. The blue lines denote the main body of the inverted U-shaped solar filament. The red lines denote the constraining magnetic fields while the dashed gray lines denote the open field lines.}\label{figcartoon}
\end{figure*}

Fig.\ref{figcartoon} illustrates the process of releasing constraint magnetic fields through magnetic reconnection initiated by the emergence of new magnetic flux. Initially, the inverted U-shaped filament (blue lines) is constrained by some magnetic fields (red lines, see panel (a)). Some magnetic dipoles emerge nearby the positive of the constraining magnetic field.  In one scenario, the positive polarity of the constraining field may connect with the negative polarity in the subsurface. The elevation of these serpentine magnetic fields would result in magnetic cancellation on the photosphere, producing long constraining magnetic fields (red lines, see panel (b)). In another scenario, the newly emerging magnetic fields reconnect with the constraining magnetic fields, forming long constraining magnetic fields (red lines) and shorter magnetic field lines (black lines, see panel (b)). Due to high curvature of the short magnetic fields, these shorter lines sink beneath the surface, leading to magnetic cancellation (see panel (b)).
In the subsequent step, the long constraining magnetic fields undergo reconnection with open magnetic fields (dashed gray lines, see panel (b)), resulting in the formation of the twisted open magnetic fields (yellow lines) and short magnetic field lines (pink lines, see panel (c)), consistent with newly forming fibrils marked by the yellow and blue dotted lines in Fig.\ref{fig3} (a3) during the second jet-like activity. These newly forming open magnetic fields, along with some already existing open magnetic fields, are released involving inner magnetic reconnection. When the twist within the open magnetic field is released, untwisted motions become observable in observations. These above episodes are consistent with the observation of two jet-like activities before eruption. As more constraining magnetic fields are progressively released, the inverted U-shaped filament becomes constrained by fewer magnetic fields (see panel (d)). Once the constraining magnetic fields can no longer maintain stability for the inverted U-shaped filament, the twisted filament would be likely to erupt.

The magnetic structure of the filament has been under debate for several decades, a flux rope \citep{kup74,aul98} or sheared arcade \citep{kip57,mal83}. Analyzing 571 filaments observed by SDO/AIA, \cite{ouy17} found that 89\% of the filaments are supported by flux ropes and 11\% are by sheared arcades. In our study, the distinguishing untwisted rotation could be found in both two jet-like activities, while untwist number released by the second activity could be more than 2 turns. On the other hand, according to the NLFFF extrapolation, a twisted structure representing the filament had been also obtained, and the twist number could be about 2.4 turns (see Fig.\ref{fig10} (d)). Furthermore, the eruptive filament showed a rotation motion during its eruption (see Fig.\ref{fig8}). These clues suggest that this active-region filament was a highly twisted flux rope rather than a sheared arcade, which is consistent with the previous study \citep[e.g.,][]{yan15,xue17,wang19}.

Through the estimations by tracing the untwisted rotation and NLFFF model extrapolation, we could derive that the twist of the filament was at least 2 turns before its eruption, which exceeds the threshold of kink instability \citep{hoo79,hoo81,tor03}. Some authors also found that the filaments/prominences had a highly twisted magnetic structure \citep[e.g.,][]{yan14,xu20}. This manifests that the highly twisted filament can be remained and does not erupt. The reason for this phenomenon might be that the heavy gravity of dense plasma or some strong constraint magnetic fields hinders the filament from erupting. In our study, we suspect that many filamentary fibrils over the filament and the magnetic fields rooting in positive polarity marked by the black box in Fig.\ref{fig2} (c2)  (represented by red lines in Fig.\ref{fig5} ) might be a key role to prevent the highly twisted filament from erupting.

The triggering mechanism of filament eruptions are crucial for understanding solar storms and forecasting interplanetary weather. On one hand, previous studies have suggested that torus instability, coupled with tether-cutting mechanism, play a key role in initiating eruptions. Tether-cutting mechanism is responsible for the slow-rise phase, while torus instability mechanism triggers the main acceleration phase \citep[e.g.,][]{cheng13,cheng20,woods2018,che18}. 
On the other hand, \cite{wyp17} proposed that magnetic breakout is a universal model for solar eruptions on different spatial scales. In the breakout model, magnetic reconnection above the main body, removing the constraint magnetic field, is involved at the onset of eruption. However, the driver or trigger for this magnetic reconnection, perhaps originating from photospheric flows or another source, remains unclear.
In our study, two jet-like precursory activities and some brightenings served as signals of magnetic reconnection, where the reconnection removed the constraint magnetic field for the filament. As the constraint magnetic fields were gradually removed, the highly twisted filament became unstable and erupted. This part is consistent with the breakout mechanism. On the other hand, we also observed that these precursory activities were associated with the emergence of newly emerging magnetic flux. In our case, the magnetic reconnection, which removed the constraint magnetic field, was caused by the newly emerging magnetic flux. This provides a more detailed episode for understanding the onset mechanism of the filament eruption.
These two jet-like activities not only eliminated some dense plasma pressing the filament but also released the constraining magnetic field around the filament, thereby increasing the gradient of magnetic pressure. In summary, these precursor activities created favorable conditions for the eruption of the highly twisted filament by removing the constraint magnetic field and some dense plasma.

\begin{acknowledgements}
We appreciate the referee's careful reading of the manuscript and many constructive comments, which helped greatly in improving the paper. We thank Dr Dong Li at Purple Mountain Observatory and Dr Hechao Chen at Yunnan University for their discussions and valuable comments.  SDO is a mission of NASA's Living With a Star Program. The authors are indebted to the SDO, STEREO, and NVST teams for providing the data. This work is supported by the National Key R\&D Program of China (2019YFA0405000), the Strategic Priority Research Program of the Chinese Academy of Sciences, Grant No. XDB0560000, the National Science Foundation of China (NSFC) under grant numbers 12003064, 12325303, 11973084, 12203020, 12203097, 12273110, 12003068, the Yunnan Key Laboratory of Solar Physics and Space Science (202205AG070009), the Yunnan Science Foundation of China under number 202301AT070347, 202201AT070194, 202001AU070077, Yunnan Science Foundation for Distinguished Young Scholars No. 202001AV070004, the grant associated with a project of the Group for Innovation of Yunnan province.
\end{acknowledgements}

\end{document}